# Solution of the Specific Model of Five-Body Problem to Investigate the Effective Alpha-Nucleon Interaction in a Partial-wave Analysis

E. Ahmadi Pouya*[1] & A. A. Rajabi[2]

Physics Department, Shahrood University of Technology, P. O. Box 3619995161-316, Shahrood, Iran

**Abstract**

In this paper, we have solved a simple specific model of the five-body problem in the framework of the Yakubovsky equations, restricted to the configurations of the alpha-nucleon types only, to investigate the effective interaction between an inert alpha-particle and a neutron. In general case, the Yakubovsky scheme for the solution of the five-body system leads to a set of four coupled equations related to four independent configurations, which can be restricted to two coupled ones, to describe the effective alpha-nucleon structure model, namely an inert four-body alpha-core and a nucleon. Hence, in such a model, the other configurations will not be taken into account. To calculate the binding energies of the five-body system in the model of alpha-nucleon structure, the two coupled equations are represented in the momentum space on the basis of the Jacobi momenta. After an explicit evaluation of the two coupled integral equations in a partial-wave analysis, the obtained equations are the starting point for a numerical calculation as an eigenvalue equation form, using typical iteration method. In the first step to the calculations, *i.e.* applying some spin-independent potential models, some obtained binding energy differences between the four-body as an alpha-particle and the five-body as an alpha-nucleon systems suggest that a simple effective interaction between an inert alpha-particle and a nucleon is attractive and of about 13 *MeV*. In addition, the represented binding energy results with respect to the regarded spin-independent potentials are in fair agreement with the obtained results from other methods.

## Introduction

The subject of effective alpha-nucleon ($\alpha N$) interaction plays an important role in nuclear structure of few-body problems that an entire understanding of this interaction is interesting and necessary. Also, the investigation of light nuclei and the study of the identity of the governing effective interactions, in addition to the specific properties of the bound and scattering states, are very interesting and relevant topics in nuclear few-body systems, as well as the atomic community. The main interest in the few-body problems are finding an accurate solution for the systems, as well as looking for unknown interactions governing on these systems. To this regard, the investigation of few-nucleon bound systems interacting via simple and realistic interactions has particularly been always in the center of interest and the description of light nuclei, and the effective $\alpha N$ interactions especially require well-established methods to the solution of the non-relativistic Schrodinger equation, in addition to the description of the relevant models of such interactions.

In the past few decades, considerable efforts have been made to investigate the effective $\alpha N$ interaction and applied to the exploration of the structure of light and heavy nuclei with this interaction, such as multichannel $\alpha N$ and $\alpha\alpha$ interactions [1], bound-state properties of the $^6$He and $^6$Li in a 3-body model, with an investigation of the $\alpha N$ interactions [2], interactions of $\alpha N$ in an elastic scattering [3], a survey of the $\alpha N$ interaction [4], peripheral $\alpha N$ scattering with $NN$ potential [5]. Also, significant attempts have been made to obtain accurate ground-state properties of the few-nucleon systems even for $A > 4$ with simple and realistic potentials, namely Stochastic Variational Monte Carlo (SVM) method [6] which, still used simplified forces, without realistic nuclear interactions and the Nonsymmetrized Hyperspherical Harmonics (HH) approach [7] appears to be quite promising to deal with permutational-symmetry breaking terms in the Hamiltonian. The HH calculational scheme is usually based on the partial-wave (PW) representation. The SVM, however, is performed directly using position vectors in the configuration space. All these methods have proved to be of great accuracy and they have been tested using different benchmarks. The successful outcomes in [6, 7] suggest that a direct study of the five-body systems and beyond is now accessible on today's computers with high computational speed. Now, it is desirable a direct treatment of the five-body problem for the model of effective alpha-nucleon ($\alpha - N$) structure, to investigate the effective interaction between an inert alpha-particle and an attractive nucleon ($\alpha N$). Therefore, in order to solve the five-body systems, we felt that an accurate and old reliable method, solution within the Yakubovsky scheme, is now interesting.

[1] E.Ahmady.ph@ut.ac.ir
[2] A.A.Rajabi@shahroodut.ac.ir

Now, after the experiences with 4- and 6-body bound systems within the Yakubovsky scheme, in a typical PW analysis [8] and a three-dimensional formalism [9] that the technical expertise has been developed and the very strong increase of computational power just recently achieved allow to study the five-body problems in the framework of the Yakubovsky equations for the model of effective $\alpha - N$ structure, to investigate the effective $\alpha N$ interaction. It is worthwhile to mention that a realistic five-nucleon problem is not allowed for a bound state. However, in order to investigate the effective interactions between the two particles, namely alpha-particle and an attractive nucleon, we consider the five-body problem for the model of effective $\alpha - N$ structure as a bound system. Also, an objection to the use of simple phenomenological potentials for $\alpha N$ scattering arises from the fact that these potentials allow a bound state, for the 5-body system which is forbidden by the exclusion principle. Therefore, in order to calculate the effective $\alpha N$ interaction in the special specific $\alpha - N$ structure of the five-body model system, we study the Yakubovsky scheme, extending the applications to systems with $A = 5$ and in order to calculate the binding energy results, we evaluate the coupled equations in momentum space based on a PW representation. Next, we have developed a particular representation of the high dimension eigenvalue matrix, which is systematic with respect to the number of components and well suited for a numerical implementation. In pursuit of this goal, we investigate the convergence of the eigenvalue of the Yakubovsky kernel with respect to the number of grid points and calculate the expectation value of the Hamiltonian operator, which is systematic with respect to the number of components and well suited for a numerical implementation.

This paper is provided as follows. In sect. I, the Yakubovsky formalism to the five-body problem using the standard notation [10] is explicitly derived. In addition, the identity of the particles is added which leads to a set of four coupled equations related to 4 different sequential sub-clusters of 5 particles. In sect. II, corresponding Jacobi coordinates of each Yakubovsky component is defined and the relevant configurations are selected to approximate the effective $\alpha - N$ structure. By these selections, a set of four coupled equations leads to two coupled ones and the irrelevant components will not be taken in to account. In sect. III, the integral representation of each wave function (WF) component is represented by introducing the PW basis states based on Jacobi momenta. Also, we describe details for numerical techniques which are considered useful for a numerical performance. In sect. IV, in order to compare and discuss our obtained results for binding energies of the four-body in the model of alpha-particle and the five-body system in the model of effective $\alpha - N$ structure, and also describe the effective $\alpha N$ interaction, the binding energy results are represented to the tables that are listed with that obtained from other methods. In addition, in order to test our calculations, we investigate the convergence of the eigenvalue of the Yakubovsky kernel with respect to the number of grid points and calculate the expectation value of the Hamiltonian operator. Finally, the conclusions are provided in sect. V.

## I.     The Five-Body Yakubovsky Formalism

In the five-body system there are ten different two-body forces, or ten different cluster decompositions ($a_4$) having 4 clusters. They are labeled by the only two-body in cluster $a_4$ they contain, e.g. $a_4 = 12 \equiv 12 + 3 + 4 + 5$. To the solution of a typical five-body bound system in Yakubovsky scheme using the sub-cluster notation [10], the idea is to first sum up the pair forces in each 4-body fragment ($a_4$), in a second step among all 3-body fragments ($a_3$), and then in a third step among all 2-body fragments ($a_2$). We work out that formalism ending with two-body sub-clusters in the spirit of the usually used approximate an effective $\alpha - N$ structure model. To this end, we start with the non-relativistic Schrödinger equation for the five-body system, as follows

$$\left( H_0 + \sum_{a_4} V_{a_4} \right) \Phi = E \, \Phi, \tag{1.1}$$

where $H_0$ stands for the free Hamiltonian operator of the five-body system, and $\sum_{a_4} V_{a_4} \equiv V_{12} + \cdots + V_{45}$, is the summation of the all 2-body interactions having ten terms. According to the Yakubovsky scheme, the Eq. (1.1) is rewritten into an integral equation

$$\Phi = G_0 \sum_{a_4} V_{a_4} \Phi, \tag{1.2}$$

where $G_0$ is the five-body free Green's function operator and in the case of scattering states we have $G_0 = [E - H_0 \pm i\varepsilon]^{-1}$. In investigating the bound states, there is no $i\varepsilon$ needed since $E < 0$. The first step is the summation of each pair force to infinite order, so we can define $\Phi = \sum_{a_4} \varphi_{a_4} \equiv \varphi_{12} + \cdots + \varphi_{45}$ for total WF where we have $\varphi_{a_4} \equiv G_0 V_{a_4} \Phi$. By inserting that decomposition for $\Phi$ on the right hand side, we have

$$\varphi_{a_4} \equiv G_0 V_{a_4} \varphi_{a_4} + G_0 V_{a_4} \sum_{b_4} \bar{\delta}_{a_4 b_4} \varphi_{b_4}. \tag{1.3}$$

The first term is related to a renewed interaction $V_{a_4}$, whereas in the second term the next interaction $V_{b_4} \neq V_{a_4}$ and we use the anti-delta function as $\bar{\delta}_{a_4 b_4} = 1 - \delta_{a_4 b_4}$. The Eq. (1.3) can be packed by using the Faddeev-like equation

$$\varphi_{a_4} \equiv G_0 t_{a_4} \sum_{b_4} \bar{\delta}_{a_4 b_4} \varphi_{b_4}, \tag{1.4}$$

where $t_{a_4}$ is a two-body $t$-matrix operator that obeys the Lippmann-Schwinger equation as $t_{a_4} = V_{a_4} + V_{a_4} G_0 t_{a_4}$. Next, we can describe main sub-clusters with definition of new components, as follows

$$\varphi_{a_4 a_3} = G_0 t_{a_4} \sum_{b_4 \subset a_3} \bar{\delta}_{a_4 b_4} \varphi_{b_4}, \tag{1.5}$$

where $(a_3)$ refers to any 3-body fragment containing the pair $(a_4)$ and the sum runs over pairs $b_4 \subset a_3$. Next, we have $\varphi_{a_4} = \sum_{a_4 \subset a_3} \varphi_{a_4 a_3}$ and this relation is used to obtain a closed set of equations for $\varphi_{a_4 a_3}$,

$$\varphi_{a_4 a_3} = G_0 t_{a_4} \sum_{b_4 \subset a_3} \bar{\delta}_{a_4 b_4} \sum_{b_4 \subset b_3} \varphi_{b_4 b_3}, \tag{1.6}$$

here we separate now the components $\varphi_{a_4 a_3}$ for a given $(a_3)$ from the rest

$$\varphi_{a_4 a_3} - G_0 t_{a_4} \sum_{b_4 \subset a_3} \bar{\delta}_{a_4 b_4} \varphi_{b_4 a_3} = G_0 t_{a_4} \sum_{b_4 \subset a_3} \bar{\delta}_{a_4 b_4} \sum_{b_4 \subset b_3} \bar{\delta}_{a_3 b_3} \varphi_{b_4 b_3}, \tag{1.7}$$

defining for a fixed $(a_3)$ the column vectors $\varphi^{a_3}$ and $\varphi^{(a_3)}$ with the components $(\varphi^{a_3})_{a_4} = \varphi_{a_4,a_3}$ and $(\varphi^{(a_3)})_{a_4} = \sum_{b_4 \subset a_3} \bar{\delta}_{a_3 b_4} \varphi_{b_4 b_3}$, respectively, and introducing the matrix $M^{a_3}$ with the elements $M^{a_3} \equiv t_{a_4} \bar{\delta}_{a_4 b_4}$ Eq. (1.7) leads to

$$\varphi^{a_3} = (1 - G_0 M^{a_3})^{-1} G_0 M^{a_3} \varphi^{(a_3)} \equiv G_0 \mathcal{T}^{a_3} \varphi^{(a_3)}. \tag{1.8}$$

It is well known that we achieve a Lippmann-Schwinger like equation in the above Faddeev like equation as $\mathcal{T}^{a_3} = M^{a_3} + M^{a_3} G_0 \mathcal{T}^{a_3}$. In the primal explicit notation and Eq. (1.8) leads to

$$\varphi_{a_4 a_3} = G_0 \sum_{b_4 \subset a_3} \mathcal{T}^{a_3}_{a_4 b_4} (\varphi^{a_3})_{a_4} = G_0 \sum_{b_4 \subset a_3} \mathcal{T}^{a_3}_{a_4 b_4} \sum_{b_4 \subset b_3} \bar{\delta}_{a_3 b_3} \varphi_{b_4 b_3}, \tag{1.9}$$

where

$$\mathcal{T}^{a_3}_{a_4 b_4} = t_{a_4} \bar{\delta}_{a_4 b_4} + G_0 \sum_{c_4 \subset a_3} t_{a_4} \bar{\delta}_{a_4 c_4} \mathcal{T}^{a_3}_{c_4 b_4}. \tag{1.10}$$

We note that there are two types of $\mathcal{T}$-matrix. For $(a_3)$ of the type $123 + 4 + 5$; $\mathcal{T}^{a_3}_{a_4 b_4}$ is a $3 \times 3$ matrix and for $(a_3)$ of the type $12 + 34 + 5$; $\mathcal{T}^{a_3}_{a_4 b_4}$ is a $2 \times 2$ matrix. Next, more decompose the right hand side of Eq. (1.9) according to 2-body fragments, are given as

$$\varphi_{a_4,a_3}^{a_2} = G_0 \sum_{\substack{b_4 \subset a_3}} \mathcal{T}_{a_4 b_4}^{a_3} \sum_{\substack{b_4 \subset b_3 \\ b_3 \subset a_2}} \bar{\delta}_{a_3 b_3} \, \varphi_{b_4 b_3}. \tag{1.11}$$

We remind that for 2-body fragments $a_2 = 1234 + 5$ type, there are 18 pairs of $(a_4)$, $(a_3)$ and for 2-body fragments $a_2 = 123 + 45$ type, there are 6 pairs of $(a_4)$, $(a_3)$. This defines the dimensions of the different $\mathcal{T}$-matrix. In the following the single particles in sub clusters 4-, 3- and 2-body fragments, *i.e.* $(a_4)$, $(a_3)$ and $(a_2)$ respectively, will no longer be displayed.

Next, we implement the identity of the particles which leads to a set of four coupled equations related to four different structures of 5 particles. After implementing the identity of the particles, in Appendix **A**, we end up with four independent components: $\varphi_{12;123}^{1234}$, $\varphi_{12;12+34}^{1234}$, $\varphi_{12,123}^{123+45}$ and $\left(\varphi_{12,12+34}^{125+34} + \varphi_{12,12+34}^{12+345}\right)$ coupled in the equations, (A.35), (A.37) and (A.39). The linear and final form of the Yakubovsky coupled equations for a general model of the five-body system yields

$$\varphi_{12;123}^{1234} = G_0 \mathcal{T}^{123}\left((P_{34}P_{45} - P_{34})\varphi_{12;123}^{1234} - P_{34}\varphi_{12,123}^{123+45} + \left(\varphi_{12,12+34}^{125+34} + \varphi_{12,12+34}^{12+345}\right) + \varphi_{12;12+34}^{1234}\right), \tag{1.12}$$

$$\varphi_{12;12+34}^{1234} = G_0 \mathcal{T}^{12+34}\left((1 - P_{34})\left((1 - P_{45})\varphi_{12;123}^{1234}\right) + (1 - P_{34})\varphi_{12,123}^{123+45}\right), \tag{1.13}$$

$$\varphi_{12,123}^{123+45} = G_0 \mathcal{T}^{123}(-P_{35})\left(\left(\varphi_{12,12+34}^{125+34} + \varphi_{12,12+34}^{12+345}\right) + \varphi_{12;12+34}^{1234}\right), \tag{1.14}$$

$$\varphi_{12,12+34}^{125+34} + \varphi_{12,12+34}^{12+345} =$$
$$G_0 \mathcal{T}^{12+34}\left((-P_{35}-P_{45})\left(\varphi_{12,12+34}^{125+34} + \varphi_{12,12+34}^{12+345}\right) - P_{45}(1-P_{34})\,\varphi_{12;12+34}^{1234} - P_{35}\left((1-P_{34})\varphi_{12;123}^{1234} + \varphi_{12,123}^{123+45}\right)\right). \tag{1.15}$$

In the next step we describe the configuration of each independent component, and select the specific configurations that describe the five-body system in the model of effective $\alpha - N$ structure.

## II. Coupled Equations of the effective alpha-nucleon structure

Regarding the sub cluster underlying the four components only two of them, that is $\varphi_{12;123}^{1234}$ and $\varphi_{12;12+34}^{1234}$, are related to the very approximative effective two-body configuration of $\alpha - N$ model, where the alpha-particle and attractive nucleon approximation is valid, according to the first two configurations in Fig. 1. The component $\varphi_{12,123}^{123+45}$ refers to an inert 3-body together with a 2-body sub clustering. The linear combinations $\left(\varphi_{12,12+34}^{125+34} + \varphi_{12,12+34}^{12+345}\right)$, refers again to an inert 3-body together with a 2-body sub clustering, where the underlying fragmentation related to 2-body fragments differs from $\varphi_{12,123}^{123+45}$ (See Fig. 1).

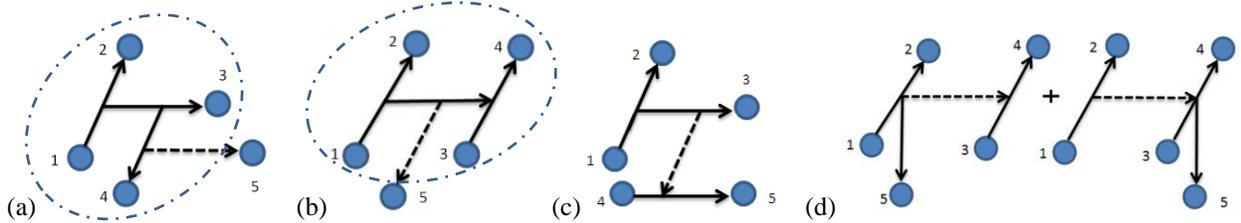

Fig. 1: Diagramatic configuration of the four independent components of 5-body system, in Jacobi coordinates. (a), (b), (c) and (d) are respectively the configurations of the components $\varphi_{12;123}^{1234}$, $\varphi_{12;12+34}^{1234}$, $\varphi_{12,123}^{123+45}$ and $\left(\varphi_{12,12+34}^{125+34} + \varphi_{12,12+34}^{12+345}\right)$. In the bag the alpha-*core* plays as a 4-body subsystem.

Now, here we explain that why we choice some specific components, and which components is related to the effective $\alpha - N$ structure. It is worthwhile to mention that for full solution of the 5-body system, in general model we need modern super-computers with grid parallel organized and we must consider the all configurations. But in this project, we are interested to study the 5-body system for the specific model of effective $\alpha - N$ structure. Therefore, according to the above discussions, we choose the two first relevant configurations, and forever the other configurations will not be taken into account in the specific $\alpha - N$ structure. Also, according to Fig. 1, the effective interaction of alpha-particle is governor and concealed in the remained components (See two first configurations in Fig. 1 and compare them with Fig. 1 and 2 in Ref. [11]). Therefore, for approximating effective $\alpha - N$ structure, we selected $\varphi_{12;123}^{1234} \equiv K$ and $\varphi_{12;12+34}^{1234} \equiv H$. By these considerations, the corresponding two first coupled equations, namely Eq. (1.12) and Eq. (1.13) leads to

$$K = G_0 \mathcal{T}^{123}((P_{34} P_{45} - P_{34}) K + H), \qquad (2.1)$$

$$H = G_0 \mathcal{T}^{12+34}(1 - P_{45} - P_{34} + P_{34} P_{45}) K. \qquad (2.2)$$

It is well-known that such a nuclear system should be treated in the fermionic approaches, *i.e.* the five-body total WF follows $\Phi = -P_{ij}\Phi$, and the Pauli principle is taken into account, even for spinless particles. But here, as a simplification, we switch off spin and isospin degrees of freedom and we study the effective five-body system in $L = 1$ states as spinless particles, (See Appendix **B**). According to the above-mentioned two first specific configurations, Fig.1, after removing the interaction of the fifth nucleon in the above-mentioned coupled equations, namely $P_{45} \equiv 0$, the five-body system leads to a typical four-body problem [11], as follows

$$K = - G_0 \mathcal{T}^{123} P_{34} K + G_0 \mathcal{T}^{123} H, \qquad (2.3)$$

$$H = G_0 \mathcal{T}^{12+34} (1 - P_{34}) K, \qquad (2.4)$$

such reduction, confirms that extending the Yakubovsky formulations for specific model of the five-body system, is resonable approximation to describe the effective $\alpha - N$ strutuer as a five-body system. Therefore, in the calculation step, for binding energies of specific five-body system, we can typically calculate the four-body binding energies to the comparison.

## III. Numerical Implementation

In this step, in order to implement the numerical techniques, the two coupled equations, Eq. (2.1) and Eq. (2.2), are represented in momentum space. Also, we introduce standard Jacobi momentum vectors on the basis of distinct configuration, according to Fig. 1, representing in Appendix **B**. Let us now represent the coupled equations, Eq. (2.1) and Eq. (2.2), to the basis states have introduced in Appendix **B**. By inserting the completeness relations, Eq. (B.9), between the permutation operators, it results

$$\langle a|K \rangle = \int a'^2 da' \int a''^2 da'' \langle a|G_0\mathcal{T}^{123}|a'\rangle\langle a'|(P_{34}P_{45} - P_{34})|a''\rangle\langle a''|K\rangle \qquad (3.1)$$
$$+ \int a'^2 da' \int b'^2 db' \langle a|G_0\mathcal{T}^{123}|a'\rangle\langle a'|b'\rangle\langle b'|H\rangle,$$

$$\langle b|H \rangle = \int b'^2 db' \int a'^2 da' \langle b|G_0\mathcal{T}^{12+34}|b'\rangle\langle b'|(1 - P_{45} - P_{34} + P_{34}P_{45})|a'\rangle\langle a'|K\rangle, \qquad (3.2)$$

the various terms appearing in the right hand side of the Eqs. (3.1) and (3.2) are explicitly evaluated in Appendix **C**. After evaluation of each term in the above-mentioned coupled integral equations in the standard PW analysis, the obtained equations are the starting point for numerical calculations as an eigenvalue equation form. In order to reduce the high dimension of the problem, first we choose an appropriate coordinate system. In this selection, the third vector $\boldsymbol{A}_3$ has been chosen parallel to $z$ −axis, the fourth vector $\boldsymbol{A}_4$ in the $x - z$ plane and the first vector $\boldsymbol{A}_1$ and second vector $\boldsymbol{A}_2$ are arbitrary in the space. Therefore, we need nine variables to uniquely specify the geometry of the four vectors $\boldsymbol{A}_i$ ($i = 1, ..., 4$) with three spherical angles and two azimuthal angles variables between them. By these considerations, the dimension of the eigenvalue problem is

$$N = N_{jac}^4 \times N_{sph}^3 \times N_{azi}^2 \times 2. \qquad (3.3)$$

The dependence on the continuous momentum and angle variables should be replaced in the numerical treatment by a dependence on certain discrete values. The high sized matrix eigenvalue equation requires an iterative solution method. We use a Lanczos-like scheme that is proved to be very efficient for nuclear few-body problems [12, 13]. This technique reduces the dimension of the eigenvalue problem to the number of iteration minus one. The evaluated coupled set Eqs. (3.1) and (3.2), in a matrix notation has the following schematic structure as an eigenvalue equation

$$\eta(E)\, \varphi(K,H) = k(E)\, \varphi(K,H), \tag{3.4}$$

where $E$ is the energy eigenvalue at which auxiliary Yakubovsky kernel eigenvalue $\eta(E)$ to be one. The Yakubovsky kernel of the linear equations $k(E)$ is energy dependent, and $\eta(E)$ is its eigenvalue with $\varphi$ as the corresponding eigenvector. In order to solve the eigenvalue equation, Eq. (3.4), we use the Gaussian quadrature grid points. The coupled equations represent a set of homogenous integral equations, which after discretization turn into a high sized matrix eigenvalue equation. Starting from an arbitrary initial $\varphi \equiv \varphi_0$ one generates by consecutive applications of $k(E)$ a sequence of amplitudes $\varphi_n$, which after orthogonalization form a basis into which $\varphi$ is expanded. It turn out that a reasonably small number of $k$-applications (of the order of 10–20) is sufficient, which leads to an algebraic eigenvalue problem of rather low dimension. Then the energy is varied such that one reaches $\eta(E) = 1$. More similar discussions can be found in refs. [8, 14].

## IV. Results
### A. Binding Energy

In this section in order to investigate the effective $\alpha N$ interaction, we present numerical results for binding energies of the five-body system in the model of effective $\alpha - N$ structure, and compare them with the four-body binding energies in the alpha-particle model, because the binding energy differences between four- and five-body systems, in such a model, refers the value of effective $\alpha N$ interaction. Bound-state results of the four- and five-body systems are shown in table 2 and 3, respectively. We also draw comparisons with results obtained by results of other methods. In order to be able to compare our calculations with results obtained by other techniques, we use the spin-independent simple potential models, as follows

I) Gauss-type Volkov potential [15]:

$$V(r) = V_R \exp[-\mu_R r^2] - V_A \exp[-\mu_A r^2]\ [MeV], \tag{4.1}$$

II) Yukawa-type Malfliet-Tjon V potential [16]

$$V(r) = V_R \frac{\exp[-\mu_R r]}{r} - V_A \frac{\exp[-\mu_A r]}{r}\ [MeV], \tag{4.3}$$

in the above-mentioned potentials, label $V_R$ and $V_A$ stand for repulsive- and attractive-part coefficients, respectively, and $\mu$ is the exchanged pion mass. The parameters of each potential are given in table 1. It is well-known such above-mentioned simple potentials applied in the calculations allow a bound state for the five-body system and they are to be expected and it is reasonably natural. In the calculations, we have used the operator form of the above potentials.

Table.1: List of the parameters of the simple potential models applied in the calculations. The potential strengths ($V_R, V_A$) are in *MeV* for Volkov and *MeV.fm* for Malfliet-Tjon V, and the range parameters, exchanged pion masses ($\mu_R, \mu_A$), are in $fm^{-2}$ for Volkov and $fm^{-1}$ for Malfliet-Tjon V.

| Potential | Type | $V_R$ | $\mu_R$ | $V_A$ | $\mu_A$ |
|---|---|---|---|---|---|
| Volko [15] | Gauss | 144.86 | 1.487 | 83.34 | 0.3906 |
| Malfliet-Tjon V [16] | Yukawa | 1458.05 | 3.11 | 578.09 | 1.55 |

For Volkov potential our calculations for four- and five-body binding energies yield the values $-30.39$ and $-44.02$ $MeV$, respectively, which as shown in table 2 are also in a good compatibility with that obtained from other calculations.

Table.2: Four- and five-body binding energies for Volkov potential in $MeV$.

| Method | $E_4$ | $E_5$ |
|---|---|---|
| HH [17,18] | $-30.420$ | $-43.032$ |
| HH [19] | $-30.406$ | $-42.383$ |
| SVM [6] | $-30.424$ | $-43.00$ |
| Present work | $-30.39$ | $-44.02$ |

Our calculations for Malfliet-Tjon V yield the value $-31.36\ MeV$ for four-body binding energy, which is in good agreement with HH [17], SVM [6] and VMC [20] results. Also our results for five-body binding energy with value $-44.30\ MeV$ are in fair compatibility with that obtained from other methods.

Table.3: Four- and five-body binding energies for Malfliet-Tjon V potential in $MeV$.

| Method | $E_4$ | $E_5$ |
|---|---|---|
| SVM [6] | $-31.360$ | $-43.48$ |
| VMC [20] | $-31.3$ | $-42.98$ |
| HH [17] | $-31.347$ | |
| Present work | $-31.36$ | $-44.30$ |

Comparison of our numerical results for binding energies with respect to the regarded spin-independent potentials are in good agreement with results of other methods in the first step calculations, and also some obtained binding energy difference between the four-body as an alpha-particle and five-body as an effective alpha-nucleon model systems suggests that an effective $\alpha N$ interaction in such a model is attractive and of about 13 $MeV$ value.

## B. Expectation Value Energy

In this section we have implemented the numerical stability of our algorithm and our representation of five-body Yakubovsky components in PW analysis. We have specially investigated the convergence of the eigenvalue of the Yakubovsky kernel with respect to the number of grid points for Jacobi momenta, azimuthal and spherical angle variables. We have also investigated the quality of our representation of the Yakubovsky components and consequently WF by calculation of the expectation value of the five-body Hamiltonian operator. We have applied the Malfliet-Tjon V potential in our investigations. In table 4 we represent the obtained eigenvalue results for five-body binding energy given in table 3 for suitable different grids. We label the number of grid points for $K$ and $H$ WFs Jacobi momenta respectively as $N_{jac}^a$ and $N_{jac}^b$, for spherical angles as $N_{sph}$ and for azimuthal angles as $N_{azi}$. As demonstrated in table 4, the calculations of the eigenvalue $\eta$ converge to the value one for $N_{jac}^a = N_{jac}^b = 20$ and $N_{sph} = N_{azi} = 14$.

Table 4. Convergence of the eigenvalue $\eta$ of Yakubovsky kernel with respect to the number of grid points in Jacobi momenta $N_{jac}^a$ and $N_{jac}^b$, spherical angles $N_{sph}$ and azimuthal angles $N_{azi}$, where $E_5 = -44.30\ MeV$.

| $N_{jac}^a$ | $N_{jac}^b$ | $N_{sph} = N_{azi}$ | $\eta$ |
|---|---|---|---|
| 10 | 10 | 14 | 0.926 |
| 14 | 10 | 14 | 0.963 |
| 16 | 14 | 14 | 0.987 |
| 20 | 16 | 14 | 0.998 |
| 20 | 20 | 14 | 1.000 |

The solution of the coupled integral equations in momentum space allows estimating numerical errors reliably. In order to demonstrate reliability of our calculations, we can evaluate the expectation value of the five-body Hamiltonian operator and compare this value to the calculated binding energy of the eigenvalue equation, Eq. (3.4), that those results are given in table 3. The expectation values of the five-body kinetic energy $\langle H_0 \rangle$, the all 2-body interaction $\langle V \rangle$ and the five-body Hamiltonian operator $\langle H \rangle$ for five-body system are given in table 5 for Malfliet-Tjon V interactions calculated in PW analysis. The little differences between the expectation value of the five-body Hamiltonian $\langle H \rangle$ and the eigenvalue energy $E_5$ shows that the results are in fairly agreement with acceptable accuracy. However, a better agreement could be reached if we considered a larger number of grid points in our calculations.

Table 5. Expectation values of the five-body kinetic energy $\langle H_0 \rangle$ and all 2-body interactions $\langle V \rangle$. Additionally, the expectation values of the five-body Hamiltonian operator $\langle H \rangle$ are compared to the binding energy results from the eigenvalue equation. All values are given in *MeV*.

| Method | $\langle H_0 \rangle$ | $\langle V \rangle$ | $\langle H \rangle$ | $E$ |
|---|---|---|---|---|
| Malfliet-Tjon V | 72.43 | −118.71 | −46.28 | −44.30 |

## V. Conclusions

The subject of alpha-nucleon interaction occupies so important a role in nuclear structure problems that makes an entire understanding of this interaction necessary. Therefore, in order to investigate the effective $\alpha N$ interaction, we have solved the coupled Yakubovsky equations for the five-body system in the model of approximating effective alpha-nucleon structure in a PW analysis that is implemented in the basis of momentum variables. To this end, we formulated the coupled equations for the spinless particles as the function of Jacobi momenta, namely the magnitudes of the momenta and the angles between them. The coupled integral equations for a bound-state calculation can be handled in PW representation and solved by a numerically reliable standard method. Our numerical results of binding energies with respect to the regarded spin-independent simple potentials are in fair agreement with the results of other methods in the first step calculations and also some obtained binding energy differences between the four-body as an alpha-particle and the five-body as an alpha-nucleon model systems suggest that an effective interaction of $\alpha N$ is attractive and occurs to about 13 *MeV*. In addition, the stability of our algorithm has been achieved with the calculation of the eigenvalue of Yakubovsky kernel, where a different number of grid points for Jacobi momenta and angle variables have been used. We have also calculated the expectation value of the five-body Hamiltonian operator. This test of calculation has been done with Malfliet-Tjon V potential and we have achieved a good compatibility between the obtained eigenvalue energy and the expectation value of the Hamiltonian operator.

It is worthwhile to mention that by including the spin effects in the implementation of the four-body system in the model of alpha-particle and five-body system in the specific model of effective $\alpha - N$ structure, both binding energy results almost will be equally improved, so, correspondingly the results of the effective $\alpha N$ interaction, will almost remain when the spin-dependent interactions are used. In addition, to the solution of the five-body system in the model of effective $\alpha - N$ structure, according to Fig. 1, only the first two relevant configurations, in terms of two first Yakubovsky components, are considered, and obviously the irrelevant configurations/components will not be taken into account, even for spin-dependent potentials. Though for a full solution of the general model of five-body bound systems, such as constituent quark models (Pentaquark) or atomic five-boson bound systems (Pentamer), the incorporation of the all components is required. This is very promising and nourishes our hope for performing calculations with spin-dependent nucleon-nucleon potential models, in a PW analysis and also 3-dimentional formalism based on Yakubovsky method.

# Appendix

## A. Implementation of the identity of the particles

We start from Eq. (1.9) choosing the case 4-body fragments $a_4 = 12$ with 3-body fragments $a_3 = 123$ one obtains

$$\varphi_{12,123} = G_0 \mathcal{T}_{12,12}^{123}\bigl(\varphi^{(123)}\bigr)_{12} + G_0 \mathcal{T}_{12,23}^{123}\bigl(\varphi^{(123)}\bigr)_{23} + G_0 \mathcal{T}_{12,31}^{123}\bigl(\varphi^{(123)}\bigr)_{31}, \tag{A.1}$$

according to the second term of Eq. (1.9)

$$\begin{aligned}\varphi_{12,123} &= G_0 \mathcal{T}_{12,12}^{123}\bigl(\varphi_{12,124} + \varphi_{12,125} + \varphi_{12,12+34} + \varphi_{12,12+35} + \varphi_{12,12+45}\bigr) \\ &+ G_0 \mathcal{T}_{12,23}^{123}\bigl(\varphi_{23,234} + \varphi_{23,235} + \varphi_{23,23+14} + \varphi_{23,23+15} + \varphi_{23,23+45}\bigr) \\ &+ G_0 \mathcal{T}_{12,31}^{123}\bigl(\varphi_{31,314} + \varphi_{31,315} + \varphi_{31,31+24} + \varphi_{31,31+25} + \varphi_{31,31+45}\bigr),\end{aligned} \tag{A.2}$$

It is easily seen, going back to the definitions Eq. (1.3) and Eq. (1.6) together with the anti-symmetry requirement for the total state wave function that

$$\varphi_{23,234} + \varphi_{23,235} + \varphi_{23,23+14} + \varphi_{23,23+15} + \varphi_{23,23+45} = P_{12}P_{23}\bigl(\varphi_{12,124} + \varphi_{12,125} + \varphi_{12,12+34} + \varphi_{12,12+35} + \varphi_{12,12+45}\bigr), \tag{A.3}$$

$$\varphi_{31,314} + \varphi_{31,315} + \varphi_{31,31+24} + \varphi_{31,31+25} + \varphi_{31,31+45} = P_{13}P_{23}\bigl(\varphi_{12,124} + \varphi_{12,125} + \varphi_{12,12+34} + \varphi_{12,12+35} + \varphi_{12,12+45}\bigr), \tag{A.4}$$

therefore, Eq. (A.2) turns into

$$\varphi_{12,123} = G_0\bigl(\mathcal{T}_{12,12}^{123} + \mathcal{T}_{12,23}^{123}P_{12}P_{23} + \mathcal{T}_{12,31}^{123}P_{13}P_{23}\bigr) \times \bigl(\varphi_{12,124} + \varphi_{12,125} + \varphi_{12,12+34} + \varphi_{12,12+35} + \varphi_{12,12+45}\bigr). \tag{A.5}$$

The coupled sets Eq. (1.10) for using $a_3 = 123$ with relations like $P_{12}P_{23}t_{13}P_{23}P_{12} = P_{13}P_{23}t_{23}P_{23}P_{13} = t_{12}$, reveals that

$$\mathcal{T}^{123} = \mathcal{T}_{12,12}^{123} + \mathcal{T}_{12,23}^{123}P_{12}P_{23} + \mathcal{T}_{12,31}^{123}P_{13}P_{23}, \tag{A.6}$$

where $\mathcal{T}^{123}$ obeys $\mathcal{T}^{123} = t_{12}P + t_{12}PG_0\mathcal{T}^{123}$ and where $P = P_{12}P_{23} + P_{13}P_{23}$. Then Eq. (A.5) simplifies to

$$\varphi_{12,123} = G_0 \mathcal{T}^{123}\bigl(\varphi_{12,124} + \varphi_{12,125} + \varphi_{12,12+34} + \varphi_{12,12+35} + \varphi_{12,12+45}\bigr). \tag{A.7}$$

Starting again from Eq. (1.9) but now for the case 4-body fragments $a_4 = 12$ with 3-body fragments $a_3 = 12 + 34$ one obtains

$$\varphi_{12,12+34} = G_0 \mathcal{T}_{12,12}^{12+34}\bigl(\varphi^{(12+34)}\bigr)_{12} + G_0 \mathcal{T}_{12,34}^{12+34}\bigl(\varphi^{(12+34)}\bigr)_{34}, \tag{A.8}$$

according to the second term of Eq. (1.9)

$$\begin{aligned}\varphi_{12,12+34} &= G_0 \mathcal{T}_{12,12}^{12+34}\bigl(\varphi_{12,123} + \varphi_{12,124} + \varphi_{12,125} + \varphi_{12,12+35} + \varphi_{12,12+45}\bigr) \\ &+ G_0 \mathcal{T}_{12,34}^{12+34}\bigl(\varphi_{34,134} + \varphi_{34,234} + \varphi_{34,345} + \varphi_{34,34+15} + \varphi_{34,34+25}\bigr),\end{aligned} \tag{A.9}$$

then we use permutation operator properties as follows

$$\varphi_{12,12+34} = G_0\bigl(\mathcal{T}_{12,12}^{12+34} + \mathcal{T}_{12,34}^{12+34}P_{13}P_{24}\bigr)\bigl(\varphi_{12,123} + \varphi_{12,124} + \varphi_{12,125} + \varphi_{12,12+35} + \varphi_{12,12+45}\bigr), \tag{A.10}$$

then defining

$$\mathcal{T}^{12+34} = \mathcal{T}_{12,12}^{12+34} + \mathcal{T}_{12,34}^{12+34}P_{13}P_{24}, \tag{A.11}$$

where $\mathcal{T}^{12+34}$ obeys the equation $\mathcal{T}^{12+34} = t_{12}\tilde{P} + t_{12}\tilde{P}G_0\mathcal{T}^{12+34}$ and where $\tilde{P} = P_{13}P_{24}$. Therefore Eq. (A.9) simplifies to

$$\varphi_{12,12+34} = G_0 \mathcal{T}^{12+34}\bigl(\varphi_{12,123} + \varphi_{12,124} + \varphi_{12,125} + \varphi_{12,12+35} + \varphi_{12,12+45}\bigr). \tag{A.12}$$

The next step is to decompose $\varphi_{12,123}$ according to Eq. (1.11). For 4-body fragments $a_4 = 12$ with 3-body fragments $a_3 = 123$ the possible 2-body fragments $(a_2)$ are $1234, 1235, 123 + 45$. Let us begin with

$$\varphi_{12,123}^{1234} = G_0 \mathcal{T}_{12,12}^{123}\bigl(\varphi_{12,124} + \varphi_{12,12+34}\bigr) + G_0 \mathcal{T}_{12,23}^{123}\bigl(\varphi_{23,234} + \varphi_{23,23+14}\bigr) + G_0 \mathcal{T}_{12,31}^{123}\bigl(\varphi_{31,134} + \varphi_{31,31+24}\bigr), \tag{A.13}$$

Since

$$\varphi_{23,234} + \varphi_{23,23+14} = P_{12}P_{23}\bigl(\varphi_{12,124} + \varphi_{12,12+34}\bigr), \tag{A.14}$$

$$\varphi_{31,134} + \varphi_{31,31+24} = P_{13}P_{23}\bigl(\varphi_{12,124} + \varphi_{12,12+34}\bigr), \tag{A.15}$$

Eq. (A.13) simplifies according to Eq. (A.6), leads to

$$\varphi_{12,123}^{1234} = G_0 \mathcal{T}^{123}\bigl(\varphi_{12,124} + \varphi_{12,12+34}\bigr), \tag{A.16}$$

similarly

$$\varphi_{12,123}^{1235} = G_0 \mathcal{T}^{123}\bigl(\varphi_{12,125} + \varphi_{12,12+35}\bigr), \tag{A.17}$$

again using symmetry properties one gets

$$\varphi_{12,123}^{123+45} = G_0 \mathcal{T}^{123}\varphi_{12,12+45}, \tag{A.18}$$

all summed up

$$\varphi_{12,123} = \varphi_{12,123}^{1234} + \varphi_{12,123}^{1235} + \varphi_{12,123}^{123+45}. \tag{A.19}$$

Which when written out agrees with Eq. (A.7). Similarly, next we decompose $\varphi_{12,12+34}$ according to Eq. (1.11). For the 4-body fragments $a_4 = 12$ with 3-body fragments $a_3 = 12 + 34$ the possible 2-body fragments ($a_2$) are $1234$, $125 + 34$, $12 + 345$, which are now regarded in turn

$$\varphi_{12,12+34}^{1234} = G_0 \mathcal{T}_{12,12}^{12+34}(\varphi_{12,123}+\varphi_{12,124}) + G_0 \mathcal{T}_{12,34}^{12+34}(\varphi_{34,234}+\varphi_{34,134}), \tag{A.20}$$

since we use $\varphi_{34,234}+\varphi_{34,134} = P_{13}P_{24}(\varphi_{12,123}+\varphi_{12,124})$. One can use Eq. (A.11) and gets

$$\varphi_{12,12+34}^{1234} = G_0 \mathcal{T}^{12+34}(\varphi_{12,123}+\varphi_{12,124}). \tag{A.21}$$

Next

$$\varphi_{12,12+34}^{125+34} = G_0 \mathcal{T}_{12,12}^{12+34}\varphi_{12,125} + G_0 \mathcal{T}_{12,34}^{12+34}(\varphi_{34,15+34}+\varphi_{34,25+34}), \tag{A.22}$$

$$\varphi_{12,12+34}^{12+345} = G_0 \mathcal{T}_{12,12}^{12+34}(\varphi_{12,12+35}+\varphi_{12,12+45}) + G_0 \mathcal{T}_{12,34}^{12+34}\varphi_{34,345}, \tag{A.23}$$

The two above amplitudes cannot be related by permutations, but their sum can be used

$$\varphi_{12,12+34}^{125+34} + \varphi_{12,12+34}^{12+345} = G_0 \mathcal{T}_{12,12}^{12+34}(\varphi_{12,125} + \varphi_{12,12+35}+\varphi_{12,12+45}) + G_0 \mathcal{T}_{12,34}^{12+34}(\varphi_{34,15+34}+\varphi_{34,25+34} + \varphi_{34,345}), \tag{A.24}$$

in the case

$$\varphi_{34,15+34}+\varphi_{34,25+34} + \varphi_{34,345} = P_{13}P_{24}(\varphi_{12,125} + \varphi_{12,12+35}+\varphi_{12,12+45}), \tag{A.25}$$

this leads to

$$\varphi_{12,12+34}^{125+34} + \varphi_{12,12+34}^{12+345} = G_0 \mathcal{T}^{12+34}(\varphi_{12,125} + \varphi_{12,12+35}+\varphi_{12,12+45}). \tag{A.26}$$

Thus, Eq. (A.20) and Eq. (A.26), summarizes to

$$\varphi_{12,12+34} = \varphi_{12,12+34}^{1234} + \varphi_{12,12+34}^{125+34} + \varphi_{12,12+34}^{12+345}, \tag{A.27}$$

Which when written out agrees with Eq. (A.12). The two amplitudes $\varphi_{12,123}^{1234}$ and $\varphi_{12,12+34}^{1234}$ expressed in Eq. (A.16) and Eq. (A.21) are connected to each other as shown now. The expression Eq. (A.19) can easily be converted to $\varphi_{12,124}$ and using in addition Eq. (A.27), with usage of Eq. (A.16), one finds

$$\varphi_{12,123}^{1234} = G_0 \mathcal{T}^{123}\left((\varphi_{12,124}^{1234} + \varphi_{12,124}^{1245} + \varphi_{12,124}^{124+35}) + (\varphi_{12,12+34}^{1234} + \varphi_{12,12+34}^{125+34} + \varphi_{12,12+34}^{12+345})\right), \tag{A.28}$$

one separates now the components $\varphi_{12,124}^{1234}$ and $\varphi_{12,12+34}^{1234}$ from the rest

$$\varphi_{12,123}^{1234} - G_0 \mathcal{T}^{123}(\varphi_{12,124}^{1234} + \varphi_{12,12+34}^{1234}) = G_0 \mathcal{T}^{123}(\varphi_{12,124}^{1245} + \varphi_{12,124}^{124+35} + \varphi_{12,12+34}^{125+34} + \varphi_{12,12+34}^{12+345}), \tag{A.29}$$

correspondingly Eq. (A.21) yields

$$\varphi_{12,12+34}^{1234} = G_0 \mathcal{T}^{12+34}\left((\varphi_{12,123}^{1234} + \varphi_{12,123}^{1235} + \varphi_{12,123}^{123+45}) + (\varphi_{12,124}^{1234} + \varphi_{12,124}^{1245} + \varphi_{12,124}^{124+35})\right), \tag{A.30}$$

again, One separates now the components $\varphi_{12,123}^{1234}$ and $\varphi_{12,124}^{1234}$ from the rest

$$\varphi_{12,12+34}^{1234} - G_0 \mathcal{T}^{12+34}(\varphi_{12,123}^{1234} + \varphi_{12,124}^{1234}) = G_0 \mathcal{T}^{12+34}(\varphi_{12,123}^{1235} + \varphi_{12,123}^{123+45} + \varphi_{12,124}^{1245} + \varphi_{12,124}^{124+35}), \tag{A.31}$$

with $\varphi_{12,124}^{1234} = -P_{34}\varphi_{12,123}^{1234}$ we can put Eq. (A.29) and Eq. (A.31) into a matrix form

$$\begin{pmatrix} \varphi_{12;123}^{1234} \\ \varphi_{12;12+34}^{1234} \end{pmatrix} - G_0 \begin{pmatrix} \mathcal{T}^{123}(-P_{34}) & \mathcal{T}^{123} \\ \mathcal{T}^{12+34}(1-P_{34}) & 0 \end{pmatrix} \begin{pmatrix} \varphi_{12;123}^{1234} \\ \varphi_{12;12+34}^{1234} \end{pmatrix} \tag{A.32}$$

$$= G_0 \begin{pmatrix} \mathcal{T}^{123}(\varphi_{12,124}^{1245} + \varphi_{12,124}^{124+35} + \varphi_{12,12+34}^{125+34} + \varphi_{12,12+34}^{12+345}) \\ \mathcal{T}^{12+34}(\varphi_{12,123}^{1235} + \varphi_{12,123}^{123+45} + \varphi_{12,124}^{1245} + \varphi_{12,124}^{124+35}) \end{pmatrix},$$

since

$$\varphi_{12,124}^{1245} = -P_{34}\varphi_{12,123}^{1235}, \varphi_{12,124}^{124+35} = -P_{34}\varphi_{12,123}^{123+45}, \tag{A.33}$$

the right-hand side of Eq. (A.32) can be factored and achieves the form as

$$\begin{pmatrix} \varphi_{12;123}^{1234} \\ \varphi_{12;12+34}^{1234} \end{pmatrix} - G_0 \begin{pmatrix} \mathcal{T}^{123}(-P_{34}) & \mathcal{T}^{123} \\ \mathcal{T}^{12+34}(1-P_{34}) & 0 \end{pmatrix} \begin{pmatrix} \varphi_{12;123}^{1234} \\ \varphi_{12;12+34}^{1234} \end{pmatrix}$$

$$= G_0 \begin{pmatrix} \mathcal{T}^{123}(-P_{34}) & \mathcal{T}^{123} \\ \mathcal{T}^{12+34}(1-P_{34}) & 0 \end{pmatrix} \begin{pmatrix} \varphi_{12,124}^{1245} + \varphi_{12,124}^{124+35} + \varphi_{12,12+34}^{125+34} + \varphi_{12,12+34}^{12+345} \\ \varphi_{12,123}^{1235} + \varphi_{12,123}^{123+45} + \varphi_{12,124}^{1245} + \varphi_{12,124}^{124+35} \end{pmatrix}, \tag{A.34}$$

the right-hand side can be reduced applying permutations and obtains final form as

$$\begin{pmatrix} \varphi_{12;123}^{1234} \\ \varphi_{12;12+34}^{1234} \end{pmatrix} = G_0 \begin{pmatrix} \mathcal{T}^{123}(-P_{34}) & \mathcal{T}^{123} \\ \mathcal{T}^{12+34}(1-P_{34}) & 0 \end{pmatrix} \left[ \begin{pmatrix} -P_{45}\varphi_{12,123}^{1234} + \varphi_{12,123}^{123+45} \\ \varphi_{12;12+34}^{125+34} + \varphi_{12;12+34}^{12+345} \end{pmatrix} + \begin{pmatrix} \varphi_{12;123}^{1234} \\ \varphi_{12;12+34}^{1234} \end{pmatrix} \right]. \tag{A.35}$$

After adequate permutation of Eq. (A.27) one obtains form Eq. (A.18) as

$$\varphi_{12,123}^{123+45} = G_0 \mathcal{T}^{123}(\varphi_{12,12+45}^{1245} + \varphi_{12,12+45}^{123+45} + \varphi_{12,12+45}^{12+345}), \tag{A.36}$$

or
$$\varphi_{12,123}^{123+45} = G_0 \mathcal{T}^{123}(-P_{35})\left(\left(\varphi_{12;12+34}^{125+34} + \varphi_{12;12+345}^{12+345}\right) + \left(\varphi_{12,12+34}^{1234}\right)\right). \tag{A.37}$$

Further Eq. (A.26) yields inserting the decomposition of the right-hand side related to
$\varphi_{12,12+34}^{125+34} + \varphi_{12,12+34}^{12+345}$

$$= G_0 \mathcal{T}^{12+34}\big(\varphi_{12,125}^{1235} + \varphi_{12,125}^{1245} + \varphi_{12,125}^{125+34} + \varphi_{12,12+35}^{1235} + \varphi_{12,12+35}^{124+35} + \varphi_{12,12+35}^{12+345} + \varphi_{12,12+45}^{1245}$$
$$+ \varphi_{12,12+45}^{123+45} + \varphi_{12,12+45}^{12+345}\big), \tag{A.38}$$

here quite a few amplitudes can be related to previous ones by permutations leading to
$\left(\varphi_{12,12+34}^{125+34} + \varphi_{12,12+34}^{12+345}\right) = G_0 \mathcal{T}^{12+34}(-P_{35}-P_{45})$

$$\times \left(\left(\varphi_{12,12+34}^{125+34} + \varphi_{12,12+34}^{12+345}\right) - P_{45}(1-P_{34})\varphi_{12,12+34}^{1234} - P_{35}\left((1-P_{34})\varphi_{12;123}^{1234} + \varphi_{12,123}^{123+45}\right)\right). \tag{A.39}$$

### B. Definition of the Jacobi momenta and partial-wave basis states

Here we display some Jacobi momenta related to the 5-body system in the case of two specific components. For $\varphi_{12;123}^{1234}$ in terms of the first configuration in Fig. 1, we choose

$$\begin{aligned}
\boldsymbol{a}_1 &= 1/2(\boldsymbol{p}_1 - \boldsymbol{p}_2), \\
\boldsymbol{a}_2 &= 1/3(2\boldsymbol{p}_3 - (\boldsymbol{p}_1 + \boldsymbol{p}_2)), \\
\boldsymbol{a}_3 &= 1/4(3\boldsymbol{p}_4 - (\boldsymbol{p}_1 + \boldsymbol{p}_2 + \boldsymbol{p}_3)), \\
\boldsymbol{a}_4 &= 1/5(4\boldsymbol{p}_5 - (\boldsymbol{p}_1 + \boldsymbol{p}_2 + \boldsymbol{p}_3 + \boldsymbol{p}_4)).
\end{aligned} \tag{B.1}$$

In the non-relativistic case we may express the kinetic energy operator by two equivalent forms. So, the kinetic energy in terms of $\boldsymbol{a}$−set Jacobi momenta, are given as

$$H_0^a = \sum_{i=1}^{5} \frac{p_i^2}{2m} \equiv \frac{a_1^2}{m} + \frac{3}{4}\frac{a_2^2}{m} + \frac{2}{3}\frac{a_3^2}{m} + \frac{5}{8}\frac{a_4^2}{m}, \tag{B.2}$$

where $p_i$ is individual particle momentum in the center of mass situation (under the condition $\sum_i \boldsymbol{p}_i = 0$) that described by relative Jacobi momenta $a_i$; ($i = 1, 2, 3, 4$). In the conventional Yakubovsky treatment the total Hamiltonian, according to Eq. (1.1), is first split into the free Hamiltonian $H_0$ and the interaction Hamiltonian (summation of all pair interactions).

Similarly, to $\varphi_{12;12+34}^{1234}$ in terms of the second configuration in Fig. 1 belongs

$$\begin{aligned}
\boldsymbol{b}_1 &= 1/2(\boldsymbol{p}_1 - \boldsymbol{p}_2), \\
\boldsymbol{b}_2 &= 1/2(\boldsymbol{p}_3 - \boldsymbol{p}_4), \\
\boldsymbol{b}_3 &= 1/2((\boldsymbol{p}_1 + \boldsymbol{p}_2) - (\boldsymbol{p}_3 + \boldsymbol{p}_4)), \\
\boldsymbol{b}_4 &= 1/5(4\boldsymbol{p}_5 - (\boldsymbol{p}_1 + \boldsymbol{p}_2 + \boldsymbol{p}_3 + \boldsymbol{p}_4)),
\end{aligned} \tag{B.3}$$

correspondingly, the kinetic energy in terms of $\boldsymbol{b}$−set Jacobi momenta, are given as

$$H_0^b = \sum_{i=1}^{5} \frac{p_i^2}{2m} \equiv \frac{b_1^2}{m} + \frac{b_2^2}{m} + \frac{1}{2}\frac{b_3^2}{m} + \frac{5}{8}\frac{b_4^2}{m}. \tag{B.4}$$

Now, we introduce the basis states corresponding to the two specific independent components. The partial-wave basis states suitable for $\varphi_{12;123}^{1234}$, are given as

$$|a\rangle \equiv |a_1 a_2 a_3 a_4; \gamma_a\rangle, \tag{B.5}$$

we represent the basis states for $\varphi_{12;12+34}^{1234}$ Jacobi momenta as

$$|b\rangle \equiv |b_1 b_2 b_3 b_4; \gamma_b\rangle, \tag{B.6}$$

we make usage of these basis states without angular momentum, spin and isospin effects, *i.e.* $\gamma_a = \gamma_b = 0$, and here we study with spinless particles. Though, in the numerical techniques we describe dependent on angular grid points by choosing the relevant coordinate systems, because total angular momentums of the 5-body system are restricted in $L = 1$ state, and the Pauli principle will be taken into accounts (See Sect. III).

Clearly, all basis states are complete in the five-body Hilbert space

$$\int A^2 DA \, |A_1 A_2 A_3 A_4\rangle\langle A_1 A_2 A_3 A_4| \equiv 1, \tag{B.9}$$

Obviously, the both basis are complete in the five-body Hilbert space. Where $A_i$ indicates each one of $a_i$ and $b_i$ magnitude of vectors, and

$$A^2 DA \equiv A_1^2 dA_1 \, A_2^2 dA_2 \, A_3^2 dA_3 \, A_4^2 dA_4, \tag{B.10}$$

also they are normalized according to

$$\langle A_1 A_2 A_3 A_4 | A'_1 A'_2 A'_3 A'_4\rangle = \frac{\delta(A_1 - A'_1)}{A_1^2} \frac{\delta(A_2 - A'_2)}{A_2^2} \frac{\delta(A_3 - A'_3)}{A_3^2} \frac{\delta(A_4 - A'_4)}{A_4^2} \tag{B.11}$$

## C. Explicit partial-wave evaluation of the coupled equations

To evaluate the coupled equations, Eq. (3.1) and (3.2), the following matrix elements need to be evaluated in a PW analysis

$$\langle a | G_0 \mathcal{T}^{123} | a'\rangle, \tag{C.1}$$

$$\langle a' | (P_{34}P_{45} + P_{34}) | a''\rangle, \tag{C.2}$$

$$\langle a' | b'\rangle, \tag{C.3}$$

$$\langle b | G_0 \mathcal{T}^{12+34} | b'\rangle, \tag{C.4}$$

$$\langle b' | (1 + P_{45} + P_{34} + P_{34}P_{45}) | a'\rangle, \tag{C.5}$$

To evaluate the first term, Eq. (C.1), we need to solve the first sub-cluster Faddeev-like equation to obtain $\mathcal{T}^{123}$ by using Pade' approximation [11], as follows

$$G_0 \mathcal{T}^{123} = G_0 t_{12} P + G_0 t_{12} P G_0 t_{12} P + G_0 t_{12} P \, G_0 t_{12} P G_0 t_{12} P + \cdots. \tag{C.6}$$

To evaluate the first term of Eq. (C.6), once more a completeness relation has to be inserted between the two-body $t$−matrix and the permutation operators

$$\langle a | G_0 t_{12} P | a'\rangle = G_0 \int a''^2 Da'' \, \langle a | t_{12} | a''\rangle \langle a'' | P | a'\rangle, \tag{C.7}$$

where

$$\langle a | t_{12} | a''\rangle = \langle a_1 | t_{12} | a''_1\rangle \langle a_2 | a''_2\rangle \langle a_3 | a''_3\rangle \langle a_4 | a''_4\rangle, \tag{C.8}$$

and

$$\langle a' | P | a''\rangle = \langle a' | P_{12} P_{23} | a''\rangle + \langle a' | P_{13} P_{23} | a''\rangle. \tag{C.9}$$

And

$$\langle a | t_{12} | a''\rangle = \langle a_1 | t_{(\epsilon)} | a''_1\rangle \frac{\delta(a_2 - a''_2)}{(a''_2)^2} \frac{\delta(a_3 - a''_3)}{(a''_3)^2} \frac{\delta(a_4 - a''_4)}{(a''_4)^2} \; ; \; \epsilon = E - \frac{3}{4}\frac{a_2^2}{m} - \frac{2}{3}\frac{a_3^2}{m} - \frac{5}{8}\frac{a_4^2}{m}, \tag{C.10}$$

where $\epsilon$ is the energy of two-body subsystem in $a$-set configuration, and

$$\langle a'' | P | a'\rangle = \frac{\delta(a''_3 - a'_3)}{(a'_3)^2} \frac{\delta(a''_4 - a'_4)}{(a'_4)^2} \int_{-1}^{1} dx_{2''2'} \frac{\delta\left[a'_1 - \left|-\frac{1}{2}a''_2 - a'_2\right|\right]}{[a'_1]^2} \frac{\delta\left[a''_1 - \left|\frac{1}{2}a'_2 + a''_2\right|\right]}{[a''_1]^2}. \tag{C.11}$$

To evaluate the term of the Eq. (C.2), there is a relation between Jacobi momenta in different chains, $(123 + 4 + 5; 12)$ and $(124 + 5 + 3; 12)$, which leads to

$$\langle a' | P_{34} P_{45} | a''\rangle = \frac{1}{2^3} \frac{\delta(a'_1 - a''_1)}{(a''_1)^2}$$
$$\times \int_{-1}^{1} da_{23} \int_{-1}^{1} da_{24} \int_{-1}^{1} da_{34} \frac{\delta\left[a'_2 - \left|\frac{1}{3}a''_2 + \frac{2}{9}a''_3 + \frac{5}{12}a''_4\right|\right]}{(a''_2)^2} \frac{\delta\left[a''_3 - \left|a''_2 + \frac{1}{12}a''_3 - \frac{5}{16}a''_4\right|\right]}{(a''_3)^2} \frac{\delta\left[a''_4 - \left|a''_3 - \frac{1}{5}a''_4\right|\right]}{(a''_4)^2}, \tag{C.12}$$

and $(123 + 4 + 5; 12)$ and $(124 + 3 + 5; 12)$,

$$\langle a' | P_{34} | a''\rangle = \frac{1}{2} \frac{\delta(a'_1 - a''_1)}{(a''_1)^2} \frac{\delta(a'_4 - a''_4)}{(a''_4)^2} \int_{-1}^{1} da_{23} \frac{\delta\left[a''_2 - \left|\frac{1}{3}a''_2 + \frac{8}{9}a''_3\right|\right]}{(a''_2)^2} \frac{\delta\left[a''_3 - \left|a''_2 - \frac{1}{3}a''_3\right|\right]}{(a''_3)^2}. \tag{C.13}$$

To evaluate the term of Eq. (C.3), there is a relation between Jacobi momenta in different chains, $(123 + 4 + 5; 12)$ and $(12 + 34 + 5; 12)$, which leads to

$$\langle a' | b'\rangle = \frac{1}{2} \frac{\delta(a'_1 - b'_1)}{(b'_1)^2} \frac{\delta(a'_4 - b'_4)}{(b'_4)^2} \int_{-1}^{1} da_{23} \frac{\delta\left[b'_2 - \left|\frac{1}{2}a''_2 - \frac{2}{3}a''_3\right|\right]}{(b'_2)^2} \frac{\delta\left[b'_3 - \left|a''_2 - \frac{2}{3}a''_3\right|\right]}{(b'_3)^2}. \tag{C.14}$$

Correspondingly, to evaluate the Eq. (C.4), we need to solve the sub-cluster Faddeev-like equation to obtain $\mathcal{T}^{12+34}$ by using Pade' approximation [11], as follows

$$G_0 \mathcal{T}^{12+34} = G_0 t_{12} \tilde{P} + G_0 t_{12} \tilde{P} \, G_0 t_{12} \tilde{P} + G_0 t_{12} \tilde{P} \, G_0 t_{12} \tilde{P} G_0 t_{12} \tilde{P} + \cdots. \tag{C.15}$$

To evaluate the first term of Eq. (C.15), once more a completeness relation has to be inserted between the two-body $t$−matrix and the permutation operators

$$\langle b | G_0 t_{12} \tilde{P} | b'\rangle = G_0 \int b''^2 Db'' \, \langle b | t_{12} | b''\rangle \langle b'' | \tilde{P} | b'\rangle, \tag{C.16}$$

where
$$\langle b|t_{12}|b''\rangle = \langle b_1|t_{12}|b''_1\rangle\langle b_2|b''_2\rangle\langle b_3|b''_3\rangle\langle b_4|b''_4\rangle, \tag{C.17}$$
and
$$\langle b''|\tilde{P}|b'\rangle = \langle b''|P_{13}P_{24}|b'\rangle. \tag{C.18}$$

The matrix elements of two-body $t$−matrix and permutation operator $\tilde{P}$ are evaluated
$$\langle b|t_{12}|b''\rangle = \langle b_1|t_{(\epsilon^*)}|b''_1\rangle \frac{\delta(b''_2 - b_2)}{(b''_2)^2}\frac{\delta(b''_3 - b_3)}{(b''_3)^2}\frac{\delta(b''_4 - b_4)}{(b''_4)^2}; \ \epsilon^* = E - \frac{b_2^2}{m} - \frac{1}{2}\frac{b_3^2}{m} - \frac{5}{8}\frac{b_4^2}{m}, \tag{C.19}$$
where $\epsilon^*$ is the energy of two-body subsystem in $b$-set configuration. To evaluate the matrix elements of permutation operator $\tilde{P}$, there is a relation between Jacobi momenta in different chains, $(12 + 34 + 5; 12)$ and $(34 + 12 + 5; 12)$,
$$\langle b''|\tilde{P}|b'\rangle = \frac{\delta(b''_1 - b'_2)}{(b'_2)^2}\frac{\delta(b''_2 - b'_1)}{(b'_1)^2}\frac{\delta(b''_3 - b'_3)}{(b'_3)^2}\frac{\delta(b''_4 - b'_4)}{(b'_4)^2}. \tag{C.20}$$

To evaluate the first term of Eq. (C.5), there is a relation between Jacobi momenta in different chains, $(12 + 34 + 5; 12)$ and $(123 + 4 + 5; 12)$
$$\langle b'|a'\rangle = \frac{1}{2}\frac{\delta(b'_1 - a'_1)}{(a'_1)^2}\frac{\delta(b'_4 - a'_4)}{(a'_4)^2}\int_{-1}^{1} db_{32'} \frac{\delta\left[b'_2 - \left|\frac{2}{3}b'_2 - \frac{2}{3}b'_3\right|\right]}{(b'_2)^2}\frac{\delta\left[b'_3 - \left|-b'_2 - \frac{1}{2}b'_3\right|\right]}{(b'_3)^2}. \tag{C.21}$$

To evaluate the second term of Eq. (C.5), there is a relation between Jacobi momenta in different chains, $(12 + 34 + 5; 12)$ and $(123 + 5 + 4; 12)$,
$$\langle b'|P_{45}|a'\rangle = \frac{1}{2^3}\frac{\delta(a'_1 - b'_1)}{(a'_1)^2}\int_{-1}^{1} db_{32'}\int_{-1}^{1} db_{42'}\int_{-1}^{1} db_{34}$$
$$\times \frac{\delta\left[a'_2 - \left|\frac{2}{3}b'_2 - \frac{2}{3}b'_3\right|\right]}{(a'_2)^2}\frac{\delta\left[a'_3 - \left|-\frac{1}{4}b'_2 - \frac{1}{8}b'_3 + \frac{3}{4}b'_4\right|\right]}{(a'_3)^2}\frac{\delta\left[a'_4 - \left|-\frac{1}{2}b'_2 - \frac{1}{4}b'_3 + \frac{1}{2}b'_4\right|\right]}{(a'_4)^2}. \tag{C.22}$$

To evaluate the third term of Eq. (C.5), there is a relation between Jacobi momenta in different chains, $(12 + 34 + 5; 12)$ and $(124 + 3 + 5; 12)$,
$$\langle b'|P_{34}|a'\rangle = \frac{1}{2}\frac{\delta(a'_1 - b'_1)}{(a'_1)^2}\frac{\delta(a'_4 - b'_4)}{(a'_4)^2}\int_{-1}^{1} db_{32'} \frac{\delta\left[a'_2 - \left|-\frac{2}{3}b'_2 - \frac{2}{3}b'_3\right|\right]}{(a'_2)^2}\frac{\delta\left[a'_3 - \left|b'_2 - \frac{1}{2}b'_3\right|\right]}{(a'_3)^2}. \tag{C.23}$$

To evaluate the fourth term of Eq. (C.5), there is a relation between Jacobi momenta in different chains, $(12 + 34 + 5; 12)$ and $(124 + 5 + 3; 12)$,
$$\langle b'|P_{34}P_{45}|a'\rangle = \frac{1}{2^3}\frac{\delta(a'_1 - b'_1)}{(a'_1)^2}\int_{-1}^{1} db_{32'}\int_{-1}^{1} db_{42'}\int_{-1}^{1} db_{34}$$
$$\frac{\delta\left[a'_2 - \left|-\frac{2}{3}b'_2 - \frac{2}{3}b'_3\right|\right]}{(a'_2)^2}\frac{\delta\left[a'_3 - \left|\frac{1}{4}b'_2 - \frac{1}{8}b'_3 + \frac{3}{4}b'_4\right|\right]}{(a'_3)^2}\frac{\delta\left[a'_4 - \left|\frac{1}{2}b'_2 - \frac{1}{4}b'_3 + \frac{1}{2}b'_4\right|\right]}{(a'_4)^2}. \tag{C.24}$$

In Appendix C the quantities $a_{ij}$ ($b_{ij}$) indicates the angle variable between $\boldsymbol{a}_i$ and $\boldsymbol{a}_j$ ($\boldsymbol{b}_i$ and $\boldsymbol{b}_j$), namely $a_{ij} \equiv \cos(a_i, a_j)$ and $b_{ij} \equiv \cos(b_i, b_j)$, respectively.